\documentclass[preprint,amsmath]{revtex4}

\DeclareMathOperator{\diag}{diag}
\DeclareMathOperator{\imag}{Im}

\begin{document}

\preprint{CFTP-IST/09-18}

\title{Reconstruction of Quark Mass Matrices with Weak Basis Texture Zeroes
from~Experimental Input}
\date{March 12, 2009}
\author{D.~Emmanuel-Costa}
\email{david.costa@ist.utl.pt}
\author{C.~Sim\~oes}
\email{csimoes@cftp.ist.utl.pt}

\affiliation{
Departamento de F\'{\i}sica and Centro de F\'{\i}sica Te\'orica de
Part\'{\i}culas (CFTP),
Instituto Superior T\'ecnico, Av. Rovisco Pais, 1049-001 Lisboa, Portugal}



\begin{abstract}
All quark mass matrices with texture zeroes obtained through weak basis
transformations are confronted with the experimental data. The reconstruction of
the quark mass matrices $M_u$ and $M_d$ at the electroweak scale is performed in
a weak basis where the matrices are Hermitian and have a maximum of three
vanishing elements. The same procedure is also accomplished for the Yukawa
coupling matrices at the grand unification scale in the context of the Standard
Model and its minimal supersymmetric extension as well as of the two Higgs
doublet model. The analysis of all viable power structures on the quark Yukawa
coupling matrices that could naturally appear from a Froggatt-Nielsen mechanism
is also presented.
\end{abstract}

\maketitle

\section{Introduction}

In spite of the enormous experimental progress, the origin of the fermion mass
pattern is still a fundamental question not yet solved in Particle Physics. Many
theoretical attempts have been made in order to go beyond the Standard Model
(SM) where a flavor symmetry would predict correctly the observed mass and
mixing hierarchies. Among many proposals in the literature, flavor gauge
symmetries in the context of grand unified theories (GUT), supersymmetric or
not, are favourite candidates~\cite{models} toward a more fundamental theory.
Examples of such possibilities are Yukawa mass matrices involving texture zeroes
and/or based on a spontaneously broken Abelian symmetry~\cite{Froggatt:1978nt}.

Generically, in the context of the SM the quark mass matrices $M_u$ and $M_d$
are complex (36 real parameters) and they have to reproduce $10$ physical
parameters, namely the six observed quark masses and their mixing angles encoded
in the Cabbibo-Kobayashi-Maskawa (CKM) unitary matrix, $V$. There is still some
freedom left that can account for this redundancy, which corresponds to the
transformation that leaves the charged currents invariant, the so-called weak
basis (WB) transformation. In theories beyond the SM transformations of this
type are expected to be less general. It turns out that some weak bases seem
more natural than others when searching for an underlying flavor symmetry. Once
the more fundamental theory is broken down to the SM, WB transformations on
the light fields scramble the main properties of the flavor symmetry.

The simplest attempt at understanding the flavor structure encoded in the
fermion mass matrices is by imposing some texture zeroes on the matrix elements.
The existence of such zeroes strongly evokes a new symmetry which, when exact,
enforces the mass matrix elements to vanish~\cite{Ramond:1993kv}. In the context
of the Froggatt-Nielsen mechanism~\cite{Froggatt:1978nt} or other
approaches~\cite{Binetruy:1994ru}, a texture zero could be understood as a
suppressed matrix element. The relevant question now is whether such texture
zeroes do have physical implications~\cite{Branco:1999nb} or can simply be
removed through an appropriate WB transformation. Thus, given the matrices $M_u$
and $M_d$, it is essential to distinguish zeroes which have no physical content
in themselves among others that imply physical constraints on the parameter
space.

Taking into account the awesome improvement in the experimental determination of
the CKM matrix~\cite{Charles:2004jd}, many texture zero structures found in the
literature~\cite{Ramond:1993kv,Binetruy:1994ru} are not consistent with the
observed data. The challenges of a model that makes an attempt to reproduce the
data arise from the precise determination of the rephasing invariant angle
$\beta\equiv\arg(-V_{cd}V_{cb}^{\ast}V_{td}^{\ast}V_{tb})$, which is rather
constrained~\cite{pdg}, and
$\gamma\equiv\arg(-V_{ud}V_{ub}^{\ast}V_{cd}^{\ast}V_{cb})$. In spite of the
large experimental errors, the measurement
of~$\gamma$~\cite{Aubert:2004qm,Chao:2004mn} is determinant due to the fact that
its extraction from input data is essentially not affected by the presence of
new physics contributions to $B^0_d-\bar{B}^0_d$ and $B^0_s-\bar{B}^0_s$
mixings~\cite{Botella:2005fc}.

The aim of this paper is the reconstruction of the quark mass matrices at the
electroweak and GUT scales in weak bases where they are Hermitian and have a
maximal number of non-physical zeroes. We also address the question whether the
reconstructed matrices at GUT scale can contain more zeroes with physical
content and eventually could exhibit a Froggatt-Nielsen flavor symmetry.
Another ambitious task of our bottom-up approach is the systematic search for
allowed Hermitian quark Yukawa structures at GUT scale.

This paper is organized as follows. In Section~\ref{sec:WBT}, we show that
starting from arbitrary matrices $M_u$ and $M_d$, it is always possible to
perform a WB transformation that renders them Hermitian, with a common zero
located at the  $(1,1)$ element and a zero in the position $(1,3)$ in the down
quark sector. We then prove that only three WB zeroes in the mass matrices are
allowed. In Section~\ref{sec:numerics}, we confront the obtained quark matrices
with the present experimental data, reconstructing them at the electroweak scale
and at a high scale where the Froggatt-Nielsen mechanism can be implemented.
Finally, in the last section we draw our conclusions.

\section{Weak Basis Texture Zeroes}
\label{sec:WBT}

In this section, we review the proof that it is always possible to make a WB
transformation such that the quark mass matrices are Hermitian and have the
following form~\cite{Branco:1999nb}
\begin{equation}
M_u=\begin{pmatrix}
0&\ast&\ast\\\ast&\ast&\ast\\\ast&\ast&\ast
\end{pmatrix}\,,\quad
M_d=\begin{pmatrix}
0&\ast&0\\\ast&\ast&\ast\\0&\ast&\ast
\end{pmatrix}\,.
\label{eq:texture13}
\end{equation}
The proof is rather straightforward if one starts from the basis where the up
mass matrix, $M^0_u$, is diagonal and the down mass matrix, $M^0_d$, is
Hermitian,
\begin{equation}
\begin{aligned}
M^0_u&=\diag(m_u,m_c,m_t),\\
M^0_d&=V\diag(m_d,m_s,m_b)\,V^{\dagger}\,.
\label{eq:physbasis}
\end{aligned}
\end{equation}
Then we look for a weak basis transformation which transforms the pair of
matrices written in Eq.~(\ref{eq:physbasis}) into the pair in
Eq.~(\ref{eq:texture13}),
\begin{equation}
\begin{aligned}
M^0_u{\rightarrow}M_{u}&=\mathcal{W}^{\dagger}\,M^0_u\,\mathcal{W}\,,\\
M^0_d{\rightarrow}M_{d}&=\mathcal{W}^{\dagger}\,M^0_d\,\mathcal{W}\,,
\label{eq:WB}
\end{aligned}
\end{equation}
where $\mathcal{W}$ is a unitary matrix which, according to
Eq.~(\ref{eq:WB}) and Eq.~(\ref{eq:physbasis}), has to satisfy the following
condition:
\begin{equation}
\left(M_u\right)_{11}=\left(M_d\right)_{11}=\left(M_d\right)_{13}=0\,.
\label{eq:zeroes}
\end{equation}
In order to make clear the existence of the unitary matrix $\mathcal{W}$, we
have split the action of $\mathcal{W}$ in two factors,
\begin{equation}
\begin{aligned}
\mathcal{W}=W\,W^{\prime}\,,
\label{eq:WBdivided}
\end{aligned}
\end{equation}
and required the unitary matrix $W$ to acquire the zeroes at the
position $(1,1)$ in both quark sectors. The unitary matrix $W^{\prime}$ is then
used to get the third zero at the position $(1,3)$ in the down quark sector,
without destroying the zeroes made through the WB transformation~$W$.

Demanding zeroes at the position $(1,1)$ implies
that the WB transformation $W$ obeys the following set of
equations~\cite{Branco:1999nb}:
\begin{subequations}
\label{eq:eqsystem}
\begin{align}
\label{eq:eqsystem1}
m_u\left|W_{11}\right|^2+m_c\left|W_{21}\right|^2+m_t\left|W_{31}\right|^2=0\,
,\\
m_d\left|X_{11}\right|^2+m_s\left|X_{21}\right|^2+m_b\left|X_{31}\right|^2=0\,,
\label{eq:eqsystem2}\\
\left|W_{11}\right|^2+\left|W_{21}\right|^2+\left|W_{31}\right|^2=1
\label{eq:eqsystem3}\,,
\end{align}
\end{subequations}
where X is given by $X\equiv V^{\dagger}W$ and $\left|X_{i1}\right|^2$ for
$i=1,2,3$ is given by
\begin{equation*}
\begin{split}
\left|X_{i1}\right|^2&=\,\left|V_{1i}\right|^2\left|W_{11}\right|^2+
\left|V_{2i}\right|^2
\left|W_{21}\right|^2+\left|V_{3i}\right|^2\left|W_{31}\right|^2\,\\
&+2Re\left(V^{\ast}_{1i}W_{11}V_{2i}W^{\ast}_{21}\right)+2Re\left(V^{\ast}_
{1i}
W_{11}V_{3i}
W^{\ast}_{31}\right)\,\\
&+2Re\left(V^{\ast}_{2i}W_{21}V_{3i}W^{\ast}_{31}\right)
\label{eq:X}\,.
\end{split}
\end{equation*}
Note however that, in order to have a solution for the system of
Eqs.~(\ref{eq:eqsystem1})-(\ref{eq:eqsystem3}), one of the quark masses in each
sector must have an opposite sign to the other two. This
requirement can always be fulfilled, since the sign of a Dirac fermion mass can
be fixed through an appropriate chiral transformation. Thus, without loss of
generality one can always restrict to the case where only one mass is negative
and the other two positive. For convenience we
write the WB transformation $W$ as
\begin{equation}
\label{eq:Wdef}
\begin{aligned}
W_{11}&=\cos\rho\,\cos\eta\,e^{i\varphi_1}\,,\\
W_{21}&=\cos\rho\,\sin\eta\,e^{i\varphi_2}\,,\\
W_{31}&=\sin\rho\,e^{i\varphi_3}\,,
\end{aligned}
\end{equation}
where the angles $\rho$ and $\eta$ are chosen in the first quadrant. Note that
with this choice the unitarity constraint given in Eq.~(\ref{eq:eqsystem3}) is
automatically verified. Once the angle $\rho$ is given, the angle $\eta$ is
simply determined from Eq.~(\ref{eq:eqsystem1}) as
\begin{equation}
\label{eq:tan}
\tan^2\eta\,=\,-\frac{m_u\cos^{2}\rho+m_c\sin^{2}\rho}{m_t}\,.
\end{equation}
The interval for the angle $\rho$ varies conforming to the sign of
the up quark masses and is restricted to
\begin{subequations}
\begin{equation}
\label{eq:rho2}
\rho\in\left[\arctan\sqrt{\tfrac{m_u}{-m_c}},\,\frac{\pi}{2}\right[\,,
\end{equation}
when $m_c$ is negative and
\begin{equation}
\label{eq:rho3}
\rho\in\left[0,\,\arctan\sqrt{\tfrac{-m_u}{m_c}}\,\right]\,,
\end{equation}
\end{subequations}
when $m_u$ is negative. Finally, Eq.~(\ref{eq:eqsystem2}) yields a relation
among the phases $\varphi_1$, $\varphi_2$ and $\varphi_3$. We also remark that
there is an infinite choice of unitary matrices $W$ with the first column given
by Eqs.~(\ref{eq:Wdef}). Such indetermination can always be seen as a
redefinition of the parameters of the unitary matrix $W^{\prime}$.

The existence of the third zero at the position $(1,3)$ is assured by
the WB transformation $W^{\prime}$, which can be parametrized as
\begin{equation}
W^{\prime}=
\begin{pmatrix}
1&0&0\\0&\cos\theta&-e^{
i\varphi} \sin\theta\\0&e^
{-i\varphi}\sin\theta&\cos\theta
\end{pmatrix}\,,
\label{eq:Wp}
\end{equation}
and do not change clearly the zeroes at the entries $(1,1)$. The condition for
having the third zero,
\begin{equation}
\left(M_d\right)_{13}=0\,,
\label{eq:md13z}
\end{equation}
implies that the angle $\theta$ is
\begin{equation}
\tan\theta=\left|\frac{(W^{\dagger}M^0_dW)_{13}}{(W^{\dagger}M^0_dW)_{12}}
\right|\,,
\end{equation}
and the phase $\varphi$ is
\begin{equation}
\varphi=\arg\left[\frac{(W^{\dagger}M^0_dW)_{13}}{(W^{\dagger}M^0_dW)_{12}}
\right]\,.
\label{eq:Wpphases}
\end{equation}

Now we address the question whether it is possible to keep both quark mass
matrices Hermitian and, simultaneously, obtain additional zeroes through WB
transformations. By counting the number of parameters present in the mass
matrices of Eq.~(\ref{eq:texture13}), one gets twelve independent real
parameters, which are more than the ten physical parameters. In principle, one
could use such freedom to perform a WB transformation in order to have more than
three zeroes. Such a WB transformation is however
not possible, thus implying that the assumption of any additional zero
does have now physical implications~\cite{Branco:1999nb}. On the other hand, if
one relaxes the assumption of Hermiticity of the quark mass matrices, it can be
shown that more zeroes can be obtained through a WB transformation,
\emph{e.g.} in the so-called parallel non-Hermitian nearest neighbor
interactions basis~\cite{Branco:1988iq}.

Another question one may raise is the possibility of having the WB zeroes
located at different positions than the WB zeroes given in
Eq.~(\ref{eq:texture13}). For example, instead of having the third zero located
at $(1,3)$, it is possible to make a WB transformation on the mass matrices in
order to obtain the third zero at the position $(1,2)$, $(2,2)$, $(2,3)$ or
$(3,3)$. To get the zero at the position $(2,2)$ we can perform  WB
transformation,
\begin{equation}
\mathcal{W}^{\prime}=
\begin{pmatrix}
1&0&0\\0&\cos\varepsilon&-\sin\varepsilon\\0&\sin\varepsilon&\cos\varepsilon
\end{pmatrix}\,,
\label{eq:Wpp}
\end{equation}
with $\varepsilon$ given by
\begin{equation}
\tan\varepsilon=\frac{{M_d^{\ }}_{23} \pm
\sqrt{{M_d^{2}}_{23}-{M_d^{\ }}_{22}{M_d^{\ }}_{33} }}{{M_d^{\ }}_{33}}\,,
\label{eq:e22}
\end{equation}
after rendering $M_d$ real in Eq.~(\ref{eq:texture13}). If instead we consider
the third zero at the position $(2,3)$, the angle $\varepsilon$ is
then given by
\begin{equation}
\tan\varepsilon=\frac{{M_d^{\ }}_{33}-{M_d^{\ }}_{22}
\pm\sqrt{({M_d^{\ }}_{33}-{M_d^{\ }}_{22})^2+4{M_d^2}_{23}}}{
2{M_d^{\ }}_{23 } }\,.
\label{eq:e23}
\end{equation}
Thus, starting from the quark mass matrices given in Eq.~(\ref{eq:texture13})
one can always obtain two new pairs of mass matrices with the third zero at the
position $(2,2)$ or $(2,3)$.

The above three pairs of mass matrices share the property of having the zeroes
at the same position $(1,1)$ for both quark sectors. There is still the
possibility to move the positions of the zeroes through new WB transformations
of the form
\begin{equation}
\label{eq:P}
\begin{aligned}
M^{\prime}_u&=P_i^T\,M_u\,P_i\,,\\
M^{\prime}_d&=P_i^T\,M_d\,P_i\,,
\end{aligned}
\end{equation}
where $P_{i},\,i=1,\dots,6$ are the six real permutation matrices, isomorphic to
$S_3\,$. We conclude that one can derive through WB transformations all pairs
$M_u$ and $M_d$ (15 pairs altogether), having one zero at the same position in
the diagonal for both sectors and an additional zero in the down-quark sector.
We could easily adapt the WB transformation $W^{\prime}$ in order to reproduce
the extra zero in the up sector instead of the down-quark sector, extending
these WB zero textures to a total of 30 pairs. This procedure can also be
implemented in the leptonic sector~\cite{Branco:2007nn}.

\section{Confronting the Weak Basis Zeroes with Experimental Data}
\label{sec:numerics}

The main goal of this paper is the reconstruction of the quark mass matrices
$M_u$ and $M_d$ in a basis where they are Hermitian with zeroes at the
same positions in the diagonal and an additional zero located in $M_d$.
In this analysis we selected, among other equivalent possibilities, the mass
matrices with the common zero at the $(1,1)$ element and the additional zero at
the position $(1,3)$. This reconstruction is performed first at the electroweak
scale, $M_Z\simeq91.2$~GeV, and later at GUT scale, $\Lambda$.

\subsection{The Quark Mass Matrices Reconstructed at $M_Z$}

We reconstruct the quark mass matrices in the weak basis given in
Eq.~(\ref{eq:texture13}), starting from Eq.~(\ref{eq:physbasis}) and
implementing the procedure described in Section \ref{sec:WBT}. In order to
construct the physical basis given in Eq.~(\ref{eq:physbasis}) at $M_Z$, we
have run all the quark masses up to $M_Z$ in the~$\overline{MS}$ scheme,
\begin{align}
\label{eq:inputi}
m_u(M_Z)&=\,1.4_{-0.5}^{+0.6}\text{ MeV}\,,\\
m_d(M_Z)&=\,2.8\pm{0.7}\text{ MeV}\,,\\
m_s(M_Z)&=\,60_{-19}^{+15}\text{ MeV}\,,\\
m_c(M_Z)&=\,0.64_{-0.09}^{+0.07}\text{ GeV}\,,\\
m_b(M_Z)&=\,2.89_{-0.08}^{+0.17}\text{ GeV}\,,\\
m_t(M_Z)&=\,170.1\pm{2.3}\text{ GeV}\,,
\end{align}
using the renormalization group equations (RGE) for QCD~\cite{qcdrge} at
three loops. The input masses employed in this RGE programme are summarized in
Appendix~\ref{sec:M-CKM}. For the quark mixing matrix, we have constructed the
unitary CKM matrix by taking the following mixing angles~\cite{pdg}
\begin{align}
\label{eq:vus}
|V_{us}|&=\,0.2255\pm0.0019\,,\\
|V_{ub}|&=\,(3.93\pm0.36)\times10^{-3}\,,\\
|V_{cb}|&=\,(41.2\pm1.1)\times10^{-3}\,.
\end{align}
In order to fully build the CKM matrix one
needs a fourth parameter that we choose to be a CP violating quantity from the
unitarity triangle - a useful graphical representation of the unitarity
relation between the first and the third column of the CKM matrix. Among many
possible choices, we have considered either the
angle $\beta$~\cite{pdg},
\begin{equation}
\label{eq:s2b}
\sin2\beta=0.681\pm0.025\,,
\end{equation}
which is rather constrained, or the angle $\gamma$~\cite{pdg},
\begin{equation}
\gamma=\left(77^{+30}_{-32}\right)^{\circ}\,.
\label{eq:inputf}
\end{equation}
The details of how the reconstruction of the CKM matrix was made for these two
choices, $\beta$ and $\gamma$, are given in Appendix~\ref{sec:M-CKM}.

With the purpose of reconstructing the mass matrices $M_u$ and $M_d$ at $M_Z$ in
the WB of Eq.~(\ref{eq:texture13}), we have scanned randomly the parameter space
taking the experimental range of the input values given in
Eqs.~(\ref{eq:inputi})-(\ref{eq:inputf}). We do not assume any correlation
(\emph{e.g.}, those arising from an alignment of the unitary matrices that
diagonalize the mass matrices) among the different matrix entries, so that our
analysis is as general as possible. We have also ensured that the random input
values taken from Eqs.~(\ref{eq:vus})-(\ref{eq:inputf}) verify the
experimental constraint on the Jarlskog invariant~\cite{pdg},
\begin{equation}
J\equiv\imag(V_{us}V_{ub}^{\ast}V_{cs}^{\ast}V_{cb})=\left(3.05^{+0.19}_{-0.20}
\right)\times10^{-5}\,.
\label{eq:Jnum}
\end{equation}
In this reconstruction, we have taken into account all possible quark mass signs
and we have varied the angle $\rho$ from the parametrization defined in
Eq.~(\ref{eq:Wdef}) according to the intervals given in
Eqs.~(\ref{eq:rho2})-(\ref{eq:rho3}). We have also included the scanning of two
unphysical phases which re-phases the CKM matrix on the left. Indeed, it is the
variation of the angle $\rho$ the main contribution for the broad range in the
quark mass matrix elements. The experimentally allowed intervals for the
elements of the quark matrices $M_u$ and $M_d$ read in GeV as
\begin{subequations}
\begin{align}
\label{eq:Mu13MZ}
|M_u|=&\left(\begin{array}{c@{\quad}cc}
0 &
0.0214-10.7 &
0.0137-2.58\\
... &
0.00358-172 &
0.00362-86.5\\
... &
... &
8.87\times10^{-8}-172
\end{array}\right)\,,\\[2mm]
\label{eq:Md13MZ}
|M_d|=&\left(\begin{array}{c@{\quad}cc}
0 &
0.00959-0.322 & 0\\
... &
0.000146-3.05&
0.0452-1.56\\
0 &
... &
0.00270-3.05
\end{array}\right)\,.
\end{align}
\end{subequations}
One sees from these Hermitian mass matrices that it does not seem viable to have
a zero at the position $(1,3)$ in up sector and at the same time to be
compatible with the experimental data. Looking at Eq.~(\ref{eq:Mu13MZ}), we
conclude that the only possibility for a new zero to appear, besides the three
WB zeroes, is in the position $(3,3)$. However such a possibility is very
unlikely, since in order to be consistent with the experimental data one needs
an enormous fine-tuning of the parameters to maintain the WB zeroes and,
simultaneously, obtain a new zero at the position $(3,3)$. It is even more
difficult to have an additional physical zero in the case that the WB zero is
located at $(2,2)$ or $(2,3)$ in the down-quark sector. The situation changes
when we consider the WB zero at the position $(1,3)$ in the up sector. In such
basis, an exact fourth zero is compatible with the experimental data as it can
be seen from the allowed range of the matrices $M_u$ and $M_d$ expressed in GeV,
\begin{subequations}
\begin{align}
\label{eq:uMu13MZ}
|M_u|=&\left(\begin{array}{c@{\quad}cc}
0 &0.254-10.6 &0\\
... &165-172 & 0-24.5\\
0 & ... & 0.00100-3.04
\end{array}\right)\,,\\[2mm]
\label{eq:uMd13MZ}
|M_d|=&\left(\begin{array}{c@{\quad}cc}
0 & 0.00116-0.312 & 0.000414-0.0171\\
...&2.75-3.06 &0.0157-0.566\\
...&... &0.00190-0.137
\end{array}\right)\,.
\end{align}
\end{subequations}
We point out that the entry $(2,3)$ of the up-quark mass matrix given in
Eq.~(\ref{eq:uMu13MZ}) can be exactly zero in agreement with the experimental
data. One can see immediately that the unitary matrices that diagonalize the
Hermitian matrices given in Eqs.~(\ref{eq:uMu13MZ}) and~(\ref{eq:uMd13MZ})
correspond to large rotations, so that the smallness of the CKM mixing angles
is obtained by huge cancellations. Therefore, we conclude that at $M_Z$ scale
more than three zeroes are not compatible with the experimental data, unless we
accept large rotations in the diagonalization process. This fact has been known
in the literature~\cite{Branco:1999tw,Mei:2003jr} and one possibility
to render viable a four texture zero Ansatz is by adding an isosinglet
vector-like quark which mixes with standard quarks~\cite{Branco:2006wv}.

Finally, we remark that we have not found significant differences in the
resulting mass matrices if one changes the CKM reconstructed by taking, in
addition to $|V_{us}|$, $|V_{ub}|$ and $|V_{cb}|$, either $\sin2\beta$ or
$\sin\gamma$, despite the experimental observations for $\sin\gamma$ have more
uncertainties.

\subsection{ The Quark Yukawa Coupling Matrices Reconstructed at
GUT scale}

\begin{table*}
\caption{\label{tab:SM} The allowed range for the Yukawa matrices $h_d$ and
$h_u$ in the SM, reconstructed for three different positions of the extra WB
zero in $h_d$, namely in the positions $(1,3)$, $(2,2)$ and $(2,3)$ at
$\Lambda=10^{14}$~GeV.}
\begin{ruledtabular}
{\scriptsize
\begin{tabular}{cccc}
& ${h_d^{}}_{13}=0$ & ${h_d^{}}_{22}=0$ & ${h_d^{}}_{23}=0$\\
\hline\\
$|h_u|$
&
$
\begin{pmatrix}
0 & \lambda^{ 2.4\,-\,5.7} & \lambda^{ 3.1\,-\,6.4}\\
\lambda^{ 2.4\,-\,5.7} & \lambda^{ 0.5\,-\,3.6} & \lambda^{ 0.9\,-\,6.5}\\
\lambda^{ 3.1\,-\,6.4} & \lambda^{ 0.9\,-\,6.5} & \lambda^{ 0.5\;-\;13.4}
\end{pmatrix}
$
&
$
\begin{pmatrix}
0 & \lambda^{ 3.6\,-\,8.8} & \lambda^{ 2.4\,-\,5.0}\\
\lambda^{ 3.6\,-\,8.8} & \lambda^{ 2.5\,-\,4.8} & \lambda^{
1.5\,-\,2.6}\\
\lambda^{ 2.4\,-\,5.0} & \lambda^{ 1.5\,-\,2.6} & \lambda^{ 0.5}
\end{pmatrix}
$
&
$
\begin{pmatrix}
0 & \lambda^{ 4.2\,-\,8.9} & \lambda^{ 2.4\,-\,5.0}\\
\lambda^{ 4.2\,-\,8.9} & \lambda^{ 3.9\;-\;7.2} & \lambda^{ 2.5\,-\,3.1}\\
\lambda^{ 2.4\,-\,5.0} & \lambda^{ 2.5\,-\,3.1} & \lambda^{0.5}
\end{pmatrix}
$
\\\\
$|h_d|$
&
$
\begin{pmatrix}
0 & \lambda^{ 4.8\,-\,7.0} & 0\\
\lambda^{ 4.8\,-\,7.0} & \lambda^{ 3.3\,-\,7.2} & \lambda^{ 3.7\,-\,5.4}\\
0 & \lambda^{ 3.7\,-\,5.4} & \lambda^{ 3.3\,-\,7.7}
\end{pmatrix}
$
&
$
\begin{pmatrix}
0 & \lambda^{ 5.8\,-\,8.0} & \lambda^{ 4.8\,-\,8.4}\\
\lambda^{ 5.8\,-\,8.0} & 0 & \lambda^{ 4.5\,-\,5.1}\\
\lambda^{ 4.8\,-\,8.4} & \lambda^{ 4.5\,-\,5.1} & \lambda^{3.3}
\end{pmatrix}
$
&
$
\begin{pmatrix}
0 & \lambda^{ 6.2\,-\,7.7} & \lambda^{ 4.8\,-\,8.2}\\
\lambda^{ 6.2\,-\,7.7} & \lambda^{ 5.6\,-\,6.9} & 0\\
\lambda^{ 4.8\,-\,8.2} & 0 & \lambda^{3.3}
\end{pmatrix}
$
\\
\end{tabular}
}
\end{ruledtabular}
\end{table*}

Grand unified models seem a natural framework for implementing family symmetries
and, in this sense, the hierarchical structure of the quark masses and mixing
angles could be generated by a flavor symmetry. We examine in detail the task
of reconstructing the quark Yukawa coupling matrices at the unification scale in
weak bases where a common zero is in the diagonal for both sectors and an
additional zero is in down-quark sector.

To fully reconstruct the Yukawa coupling matrices $h_u$ and $h_d$ at GUT scale,
$\Lambda$, starting from the input values given at electroweak scale, we run the
Yukawa couplings by using their 1-loop renormalization group
equations~\cite{rge}. Note however that, for the RGE programme from low energy
to the GUT scale, we need in addition to take the charged lepton masses, we have
used their values from Particle Data Group~\cite{pdg}. This RGE evolution was
made by considering three different effective low energy models: SM, SM extended
with an extra Higgs doublet (DHM) and minimal supersymmetric extension of the SM
(MSSM). In the case of the SM, the quark Yukawa coupling matrices $h_{u}$ and
$h_{d}$ at $M_Z$ are related  to the quark mass matrices $M_{u}$ and $M_{d}$
through the following relations
\begin{equation}
\label{eq:ySM}
h_{u}(M_{Z})=\frac{M_{u}(M_{Z})}v\,,\quad
h_{d}(M_{Z})=\frac{M_{d}(M_{Z})}v\,,
\end{equation}
where $v=174.1$ GeV is the vacuum expectation value (VEV) of the Higgs doublet.
These relations change to
\begin{equation}
\label{eq:yDHM}
h_{u}(M_{Z})=\frac{M_{u}(M_{Z})}{v\sin\beta}\,,\quad
h_{d}(M_{Z})=\frac{M_{d}(M_{Z})}{v\cos\beta}\,,
\end{equation}
in the case of extending the SM with an extra Higgs doublet with opposite
hypercharge. In Eq.~(\ref{eq:yDHM}) we have parametrized the VEVs with $\beta$,
defined by the ratio $\tan\beta\equiv v_{u}/v_{d}$, and $v$, verifying
$v_u^2+v_d^2=v^2$. The quantities $v_u$ and $v_d$ are the VEVs of Higgs doublets
that couple to the up and down quark sectors, respectively.

The gauge couplings do not exactly unify in either SM or DHM cases without
further assumptions. For illustration, we assumed the unification scale at
$\Lambda=10^{14}$~GeV. We then calculated the Yukawa coupling matrices at GUT
scale $\Lambda$ by making use of the RGE~\cite{rge} and taking as initial
conditions Eqs.~(\ref{eq:Mu13MZ}) and~(\ref{eq:Md13MZ}) through the relations of
Eq.~(\ref{eq:ySM}) for the SM and Eq.~(\ref{eq:yDHM}) for the DHM.

Within the context of the MSSM, the gauge couplings measured at $M_Z$ are
consistent with a single unified coupling constant at the scale
$\Lambda\simeq2\times10^{16}$~GeV. Assuming $M_S$ the natural scale above which
the MSSM is valid, the running of the quark Yukawa couplings is done in two
steps. First, we run the Yukawa couplings $h_u$ and $h_d$ from $M_Z$ to $M_S$ by
using the RGE for the SM. Then, at the threshold $M_S$ we match the Yukawa
coupling matrices $Y_u$ and $Y_d$ of the MSSM with the Yukawa coupling matrices
$h_u$ and $h_d$ of the SM as
\begin{equation}
\label{eq:yMSSM}
Y_{u}(M_S)=\frac{h_{u}(M_S)}{\sin\beta}\,,\quad
Y_{d}(M_S)=\frac{h_{d}(M_S)}{\cos\beta}\,.
\end{equation}
In the numerical analysis the threshold scale $M_S$ was chosen to be between~1
and~10 TeV. Finally, we run the Yukawa
coupling matrices $Y_u$ and $Y_d$ from $M_S$ to
$\Lambda=2\times10^{16}$~GeV and derived the following approximate
hierarchical relations for up and down quark Yukawa couplings at GUT scale,
\begin{subequations}
\begin{align}
\label{eq:popt1}
y_t:y_c:y_u\,&\approx1:\lambda^4:\lambda^8\,,\\
\label{eq:popt2}
y_b:y_s:y_d\,&\approx1:\lambda^{2-3}:\lambda^{4-5}\,,
\end{align}
\end{subequations}
which are written in terms of powers of the Cabbibo angle, $\lambda$, fixed as
$\lambda=0.22$. The range of the powers in Eqs.~(\ref{eq:popt2}) also reflects
the fact we have scanned $\tan\beta$ from $10$ to $50$. We would like to
remark that these relations also hold for the non-supersymmetric cases (SM and
DHM).

\begin{table*}
\caption{\label{tab:DHM} The Yukawa matrices $h_u$ and $h_d$ in the context of
DHM, reconstructed for three different positions of the extra WB zero in $h_d$,
namely in the positions $(1,3)$, $(2,2)$ and $(2,3)$ at $\Lambda=10^{14}$~GeV,
and for $\tan\beta=10$ and $\tan\beta=50$.}
\begin{ruledtabular}
{\scriptsize
\begin{tabular}{ccccc}
$\tan\beta$ & & ${h_d^{}}_{13}=0$ & ${h_d^{}}_{22}=0$ & ${h_d^{}}_{23}=0$
\\\hline\\
&
$|h_u|$
&
$
\begin{pmatrix}
0 & \lambda^{ 2.4\,-\,6.1} & \lambda^{ 3.3\,-\,6.6}\\
\lambda^{ 2.4\,-\,6.1} & \lambda^{ 0.5\,-\,4.9} & \lambda^{ 0.9\,-\,7.1}\\
\lambda^{ 3.3\,-\,6.6} & \lambda^{ 0.9\,-\,7.1} & \lambda^{ 0.5\,-\,13.1}
\end{pmatrix}
$
&
$
\begin{pmatrix}
0 & \lambda^{ 3.6\,-\,8.4} & \lambda^{ 2.4\,-\,4.9}\\
\lambda^{ 3.6\,-\,8.4} & \lambda^{ 2.6\,-\,5.1} & \lambda^{ 1.6\,-\,2.7}\\
\lambda^{ 2.4\,-\,4.9} & \lambda^{ 1.6\,-\,2.7} & \lambda^{0.5}
\end{pmatrix}
$
&
$
\begin{pmatrix}
0 & \lambda^{ 4.3\,-\,9.5} & \lambda^{ 2.4\,-\,4.8}\\
\lambda^{ 4.3\,-\,9.5} & \lambda^{ 3.9\,-\,6.4} & \lambda^{ 2.6\,-\,3.3}\\
\lambda^{ 2.4\,-\,4.8} & \lambda^{ 2.6\,-\,3.3} & \lambda^{0.5}
\end{pmatrix}
$
\\[4mm]
$10$ &&&
\\[4mm]
&
$|h_d|$
&
$
\begin{pmatrix}
0 & \lambda^{ 3.4\,-\,5.6} & 0\\
\lambda^{ 3.4\,-\,5.6} & \lambda^{ 1.8\,-\,7.8} & \lambda^{ 2.2\,-\,4.0}\\
0 & \lambda^{ 2.2\,-\,4.0} & \lambda^{ 1.8\,-\,6.4}
\end{pmatrix}
$
&
$
\begin{pmatrix}
0 & \lambda^{ 4.4\,-\,6.6} & \lambda^{ 3.3\,-\,8.5}\\
\lambda^{ 4.4\,-\,6.6} & 0 & \lambda^{ 3.0\,-\,3.8}\\
\lambda^{ 3.3\,-\,8.5} & \lambda^{ 3.0\,-\,3.8} & \lambda^{ 1.8\,-\,1.9}
\end{pmatrix}
$
&
$
\begin{pmatrix}
0 & \lambda^{ 4.8\,-\,6.3} & \lambda^{ 3.4\,-\,7.5}\\
\lambda^{ 4.8\,-\,6.3} & \lambda^{ 4.3\,-\,5.7} & 0\\
\lambda^{ 3.4\,-\,7.5} & 0 & \lambda^{ 1.8\,-\,1.9}
\end{pmatrix}
$
\\[8mm]\hline\\
&
$|h_u|$
&
$
\begin{pmatrix}
0 & \lambda^{ 2.3\,-\,6.2} & \lambda^{ 3.3\,-\,6.7}\\
\lambda^{ 2.3\,-\,6.2} & \lambda^{ 0.5\,-\,6.8} & \lambda^{ 0.9\,-\,7.6}\\
\lambda^{ 3.3\,-\,6.7} & \lambda^{ 0.9\,-\,7.6} & \lambda^{ 0.5\,-\,14.9}
\end{pmatrix}
$
&
$
\begin{pmatrix}
0 & \lambda^{ 3.7\,-\,7.7} & \lambda^{ 2.4\,-\,5.0}\\
\lambda^{ 3.7\,-\,7.7} & \lambda^{ 2.7\,-\,5.5} & \lambda^{ 1.6\,-\,2.8}\\
\lambda^{ 2.4\,-\,5.0} & \lambda^{ 1.6\,-\,2.8} & \lambda^{0.5}
\end{pmatrix}
$
&
$
\begin{pmatrix}
0 & \lambda^{ 4.3\,-\,8.5} & \lambda^{ 2.3\,-\,5.0}\\
\lambda^{ 4.3\,-\,8.5} & \lambda^{ 3.9\,-\,6.8} & \lambda^{ 2.6\,-\,3.4}\\
\lambda^{ 2.3\,-\,5.0} & \lambda^{ 2.6\,-\,3.4} & \lambda^{0.5}
\end{pmatrix}
$
\\[4mm]
$50$ &&&
\\[4mm]
&
$|h_d|$
&
$
\begin{pmatrix}
0 & \lambda^{ 2.1\,-\,4.4} & 0\\
\lambda^{ 2.1\,-\,4.4} & \lambda^{ 0.5\,-\,7.5} & \lambda^{ 1.0\,-\,3.4}\\
0 & \lambda^{ 1.0\,-\,3.4} & \lambda^{ 0.5\,-\,5.2}
\end{pmatrix}
$
&
$
\begin{pmatrix}
0 & \lambda^{ 3.2\,-\,5.4} & \lambda^{ 2.1\,-\,7.3}\\
\lambda^{ 3.2\,-\,5.4} & 0 & \lambda^{ 1.8\,-\,2.6}\\
\lambda^{ 2.1\,-\,7.3} & \lambda^{ 1.8\,-\,2.6} & \lambda^{ 0.5\,-\,0.6}
\end{pmatrix}
$
&
$
\begin{pmatrix}
0 & \lambda^{ 3.6\,-\,5.6} & \lambda^{ 2.1\,-\,7.4}\\
\lambda^{ 3.6\,-\,5.6} & \lambda^{ 3.1\,-\,4.7} & 0\\
\lambda^{ 2.1\,-\,7.4} & 0 & \lambda^{ 0.5\,-\,0.6}
\end{pmatrix}
$
\\[8mm]
\end{tabular}
}
\end{ruledtabular}
\end{table*}

\begin{table*}
\caption{\label{tab:MSSM} The Yukawa matrices $Y_u$ and $Y_d$ in the context of
the MSSM, reconstructed for three different positions of the extra WB zero in
$Y_d$, namely in the positions $(1,3)$, $(2,2)$ and $(2,3)$ at
$\Lambda=2\times10^{16}$~GeV, with $M_S=1$~TeV, and for
$\tan\beta=10$ and $\tan\beta=50$.}
\begin{ruledtabular}
{\scriptsize
\begin{tabular}{c c c c c}
$\tan\beta$ & & ${Y_d^{}}_{13}=0$ & ${Y_d^{}}_{22}=0$ & ${Y_d^{}}_{23}=0$\\
\hline\\
&
$|Y_u|$
&
$
\begin{pmatrix}
0 & \lambda^{ 2.4\,-\,6.0} & \lambda^{ 3.2\,-\,6.4}\\
\lambda^{ 2.4\,-\,6.0} & \lambda^{ 0.4\,-\,4.5} & \lambda^{ 0.8\,-\,5.9}\\
\lambda^{ 3.2\,-\,6.4} & \lambda^{ 0.8\,-\,5.9} & \lambda^{ 0.4\;-\;12.4}
\end{pmatrix}
$
&
$
\begin{pmatrix}
0 & \lambda^{ 3.7\,-\,9.1} & \lambda^{ 2.4\,-\,4.8}\\
\lambda^{ 3.7\,-\,9.1} & \lambda^{ 2.6\,-\,4.7} & \lambda^{ 1.5\,-\,2.4}\\
\lambda^{ 2.4\,-\,4.8} & \lambda^{ 1.5\,-\,2.4} & \lambda^{ 0.4\,-\,0.5}
\end{pmatrix}
$
&
$
\begin{pmatrix}
0 & \lambda^{ 4.3\,-\,8.5} & \lambda^{ 2.4\,-\,4.9}\\
\lambda^{ 4.3\,-\,8.5} & \lambda^{ 3.9\,-\,6.0} & \lambda^{ 2.5\,-\,3.0}\\
\lambda^{ 2.4\,-\,4.9} & \lambda^{ 2.5\,-\,3.0} & \lambda^{0.4}
\end{pmatrix}
$
\\[4mm]
$10$ &&&
\\[4mm]
&
$|Y_d|$
&
$
\begin{pmatrix}
0 & \lambda^{ 3.5\,-\,5.7} & 0\\
\lambda^{ 3.5\,-\,5.7} & \lambda^{ 1.9\,-\,7.4} & \lambda^{ 2.3\,-\,4.1}\\
0 & \lambda^{ 2.3\,-\,4.1} & \lambda^{ 1.9\,-\,6.3}
\end{pmatrix}
$
&
$
\begin{pmatrix}
0 & \lambda^{ 4.6\,-\,6.7} & \lambda^{ 3.5\,-\,7.7}\\
\lambda^{ 4.6\,-\,6.7} & 0 & \lambda^{ 3.1\,-\,3.6}\\
\lambda^{ 3.5\,-\,7.7} & \lambda^{ 3.1\,-\,3.6} & \lambda^{1.9}
\end{pmatrix}
$
&
$
\begin{pmatrix}
0 & \lambda^{ 5.0\,-\,6.4} & \lambda^{ 3.5\,-\,7.2}\\
\lambda^{ 5.0\,-\,6.4} & \lambda^{ 4.4\,-\,5.4} & 0\\
\lambda^{ 3.5\,-\,7.2} & 0 & \lambda^{1.9}
\end{pmatrix}
$
\\\\\hline\\
&
$|Y_u|$
&
$
\begin{pmatrix}
0 & \lambda^{ 2.3\,-\,5.8} & \lambda^{ 3.2\,-\,6.5}\\
\lambda^{ 2.3\,-\,5.8} & \lambda^{ 0.3\,-\,3.7} & \lambda^{ 0.8\,-\,7.2}\\
\lambda^{ 3.2\,-\,6.5} & \lambda^{ 0.8\,-\,7.2} & \lambda^{ 0.3\,-\,14.5}
\end{pmatrix}
$
&
$
\begin{pmatrix}
0 & \lambda^{ 3.7\,-\,8.3} & \lambda^{ 2.3\,-\,4.7}\\
\lambda^{ 3.7\,-\,8.3} & \lambda^{ 2.6\,-\,5.1} & \lambda^{ 1.5\,-\,2.6}\\
\lambda^{ 2.3\,-\,4.7} & \lambda^{ 1.5\,-\,2.6} & \lambda^{ 0.3\,-\,0.4}
\end{pmatrix}
$
&
$
\begin{pmatrix}
0 & \lambda^{ 4.3\,-\,8.2} & \lambda^{ 2.3\,-\,4.8}\\
\lambda^{ 4.3\,-\,8.2} & \lambda^{ 3.9\,-\,6.4} & \lambda^{ 2.5\,-\,3.2}\\
\lambda^{ 2.3\,-\,4.8} & \lambda^{ 2.5\,-\,3.2} & \lambda^{ 0.3\,-\,0.4}
\end{pmatrix}
$
\\[4mm]
$50$ &&&
\\[4mm]
&
$|Y_d|$
&
$
\begin{pmatrix}
0 & \lambda^{ 2.1\,-\,4.5} & 0\\
\lambda^{ 2.1\,-\,4.5} & \lambda^{ 0.5\,-\,5.4} & \lambda^{ 0.9\,-\,2.8}\\
0 & \lambda^{ 0.9\,-\,2.8} & \lambda^{ 0.5\,-\,5.2}
\end{pmatrix}
$
&
$
\begin{pmatrix}
0 & \lambda^{ 3.3\,-\,5.5} & \lambda^{ 2.1\,-\,6.3}\\
\lambda^{ 3.3\,-\,5.5} & 0 & \lambda^{ 1.8\,-\,2.5}\\
\lambda^{ 2.1\,-\,6.3} & \lambda^{ 1.8\,-\,2.5} & \lambda^{ 0.5\,-\,0.6}
\end{pmatrix}
$
&
$
\begin{pmatrix}
0 & \lambda^{ 3.7\,-\,5.2} & \lambda^{ 2.1\,-\,6.2}\\
\lambda^{ 3.7\,-\,5.2} & \lambda^{ 3.1\,-\,4.6} & 0\\
\lambda^{ 2.1\,-\,6.2} & 0 & \lambda^{ 0.5\,-\,0.6}
\end{pmatrix}
$
\\[8mm]
\end{tabular}
}
\end{ruledtabular}
\end{table*}

The reconstructed Yukawa coupling matrices $h_u$ and $h_d$ are presented in
Table \ref{tab:SM} for the SM and in Table \ref{tab:DHM} for the DHM at the
unification scale $\Lambda=10^{14}$~GeV by taking into account the scanning of
all electroweak input parameters. Rather than to list the numerical values for
the Yukawa coupling matrix elements, we have written them in terms of powers of
$\lambda$~\cite{Ramond:1993kv,Binetruy:1994ru,Branco:1999tw}, considering
different WB with zeroes at positions $(1,3)$, $(2,2)$ and $(2,3)$ in the
down-quark sector. Since the Yukawa coupling matrices $h_u$ and $h_d$ for the
DHM depend on $\tan\beta$, we have particularized in Table~\ref{tab:DHM} the
cases for $\tan\beta=10$ and $\tan\beta=50$.

For the case of MSSM, we show in Table~\ref{tab:MSSM} the Yukawa coupling
matrices $Y_u$ and $Y_d$ reconstructed at $\Lambda=2\times10^{16}$~GeV with
$M_S=1$~TeV for similar WB zeroes. Again, we write the elements of the Yukawa
coupling matrices in powers of Cabbibo angle $\lambda$ considering two values
of $\tan\beta$: 10 and 50. From the numerical calculations, we have verified
that the powers of $\lambda$ shown in Table~\ref{tab:MSSM} are not much affected
by varying the threshold scale $M_S$ from $1$~TeV to $10$~TeV. They are
however affected by the choice of $\tan\beta$, since the down-quark Yukawa
matrix increases proportionally with $\tan\beta\,$. In what concerns the
up-quark sector the variation is rather smooth and difficult to distinguish.
This property remains also valid for the case of DHM. We see from our results
that the Yukawa power structures vary along the various low effective models and
the positions of the WB zeroes. For instance, the range of the matrix element
$(3,3)$ for both sectors is really narrow for the cases when the extra WB
zero is at the position $(2,2)$ or $(2,3)$ in the down sector. This feature does
not depend much on $\tan\beta$ (DHM and MSSM).

For the case where the WB zero in the down sector is at the position $(1,3)$,
the interval of the up-quark Yukawa matrix element $(3,3)$ is large and the
matrix element could even be negligibly small. Hence, new possible zeroes with
physical implications could be searched. We have found that a new meaningful
zero at $(3,3)$ in the up sector could be acceptable for all considered models
and independently of $\tan\beta$. On the other hand, such possibility does not
seem viable in the context of a flavor symmetry, since it would imply that the
CKM matrix is derived from large cancellations of up- and down-quark left
rotation matrices, ${h_u^{}}_{33}\ll{h_u^{}}_{ij},$ for $(i,j)\neq(3,3)$. A
deeper analysis would be needed in order to relate the new texture zeroes
emerging from our results with the viable four texture zeroes found in the
literature~\cite{Ramond:1993kv,Branco:1999tw,Mei:2003jr,Du:1992iy}.

The underlying motivation to write the quark Yukawa coupling matrix elements in
terms of powers of $\lambda$ is that such power structure may lead to an insight
of the flavor content beyond the Standard Model. At this point, it is clear the
relevance of choosing the adequate weak basis where a new symmetry could appear
naturally, thus explaining the quark mass and their mixing hierarchies. On the
other hand, if a flavor symmetry is responsible for a power structure in the
quark Yukawa coupling matrices, it is natural to expect that the smallness of
CKM mixing angles should not be due to a relative fine-tuning of up- and
down-quark left rotations. Having this criterion of small mixing angles in mind,
new patterns of power structures appear at GUT scale. For the case of the SM we
obtained approximately the following structure:
\begin{equation}
|h_u|\propto\left(\begin{array}{c@{\quad}cc}
0 & \lambda^{5} & \lambda^{4}\\
\lambda^{5} & \lambda^{\frac52} & \lambda\\
\lambda^{4} & \lambda & 1
\end{array}\right)\,,\quad
|h_d|\propto\left(\begin{array}{c@{\quad}cc}
0 & \lambda^{\frac72} & 0 \\
\lambda^{\frac72} & \lambda^{\frac52} & \lambda \\
0 & \lambda & 1
\end{array}\right)\,.
\end{equation}
These results change in the case of the MSSM, we obtained for $\tan\beta=10$,
\begin{equation}
|Y_u|\propto\left(\begin{array}{c@{\quad}cc}
0 &\lambda^{5}&\lambda^{4}\\
\lambda^{5} & \lambda^{3}&\lambda^{2}\\
\lambda^{4} & \lambda^{2} & 1
\end{array}\right)\,,\quad
|Y_d|\propto\left(\begin{array}{c@{\quad}cc}
0 & \lambda^{4} & 0 \\
\lambda^{4} & \lambda^{2} & \lambda^{2} \\
0 & \lambda^{2} & 1
\end{array}\right)\,,
\end{equation}
and for $\tan\beta=50$,
\begin{equation}
|Y_u|\propto\left(\begin{array}{c@{\quad}cc}
0 &\lambda^{5}&\lambda^{4}\\
\lambda^{5} & \lambda^{3}&\lambda\\
\lambda^{4} & \lambda & 1
\end{array}\right)\,,\quad
|Y_d|\propto\left(\begin{array}{c@{\quad}cc}
0 & \lambda^{4} & 0 \\
\lambda^{4} & \lambda^{4} & \lambda^{2} \\
0 & \lambda^{2} &1
\end{array}\right)\,.
\end{equation}
These type of power structures could be naturally realized in the context of the
Froggatt-Nielsen mechanism~\cite{Froggatt:1978nt}, where the quark Yukawa
couplings arise from non-renormalizable interactions after a scalar singlet
field acquires a vacuum expectation value.

\section{Conclusions}

With the goal to search for new texture zero structures compatible with the
experimental data, we have
reconstructed the quark mass matrices at the $M_Z$ scale in the basis where the
matrices are Hermitian and have a maximum of three vanishing elements (one
common zero at the same position in the diagonal and an extra zero in either
sector). We found that it is unlikely
to have more zeroes than the WB zeroes at $M_Z$ scale. Thus, having a new zero
beyond the three WB zeroes implies physical constraints on the parameters. For
instance, a ``parallel'' structure, with zeroes located at $(1,1)$ and $(1,3)$
elements for both quark mass matrices, is not compatible with the electroweak
data, which requires $0.0137\text{ GeV} \lesssim|{M_u^{}}_{13}|
\lesssim2.58\text{ GeV}$.

In addition, we have reconstructed the quark Yukawa couplings in several weak
bases with texture zeroes at GUT scale. This was done by considering three low
energy models below the GUT scale: SM, DHM and MSSM. Having in mind a gauge
flavor symmetry (Froggatt-Nielsen mechanism) that would explain the hierarchy
of quark masses and their mixings, in Tables \ref{tab:SM}, \ref{tab:DHM} and
\ref{tab:MSSM} we presented viable power structures of the reconstructed Yukawa
coupling matrices as a function of the Cabbibo angle. We showed that if one
requires that the smallness of the CKM mixing angles is obtained through small
up- and down-quark left rotations, a new pattern of texture zeroes appears.
We have also emphasized the importance of right choice of a weak basis where
the implementation of the certain flavor symmetry naturally reveals.

\begin{acknowledgments}
We would like to thank Gustavo C. Branco and Ricardo Gonz\'alez Felipe
for fruitful discussions and reading carefully the paper.
This work was partially supported by Funda\c c\~ ao para a Ci\^encia e a
Tecnologia, Portugal, through the projects POCTI/FNU/44409/2002,
PDCT/FP/63914/2005, PDCT/FP/63912/2005 and CFTP-FCT UNIT 777, partially funded
by POCTI (FEDER) and by the Marie Curie RTN MRTNCT-2006-035505.
\end{acknowledgments}

\appendix

\section{Masses and CKM Reconstruction at $M_Z$ scale}
\label{sec:M-CKM}

In this appendix, we summarize all the input values needed at $M_Z$ scale.
First, we address the question how to run the quark masses to $M_Z$ by the QCD
RGE programme. Then we present two procedures for the reconstruction of the CKM
matrix, starting from three moduli, $|V_{us}|$, $|V_{ub}|$, $|V_{cb}|$ and
either $\sin2\beta$ or $\sin\gamma$.

To compute the quark running masses at $M_Z$ we have used the set of RGEs for
QCD~\cite{qcdrge} using a mass-independent substraction scheme, the
$\overline{MS}$ scheme. The u-, d-, s-quark masses have been estimated at a
scale $\mu\approx2$ GeV as follows~\cite{pdg}
\begin{align}
&m_u=\,2.4\pm{0.9}\text{ MeV}\,,\\
&m_d=\,4.75\pm{1.25}\text{ MeV}\,,\\
&m_s=\,104\pm{26}\text{ MeV}\,,
\end{align}
and verifying the relations
\begin{align}
&m_u/m_d=0.35-0.6,\\
&m_s/m_d=17-22,\\
&(m_u+m_d)/2=2.5-5.0\,\text{MeV},\\
&(m_s-(m_u+m_d)/2)/(m_d-m_u)=30-50\,.
\end{align}
For the heavy quark running masses we have used~\cite{pdg}
\begin{align}
m_c(m_c)&=\,1.27_{-0.11}^{+0.07}\text{ GeV}\,,\\
m_b(m_b)&=\,4.20_{-0.11}^{+0.17}\text{ GeV}\,,\\
M^{\text{\scriptsize pole}}_t&=\,171.2\pm{2.1}\text{ GeV}\,.
\end{align}

To reconstruct the CKM matrix, some parametrizations seem to be more adequate
than others, even though they have no special meaning by themselves. The best
evidence for a complex CKM mixing matrix arises from the angle
$\gamma$~\cite{Botella:2005fc}, and consequently it is a good indicator of the
presence of new physics. On the other hand, $\gamma$ is the least known angle of
the unitarity triangle and it is still limited by the statistical and
theoretical uncertainties. On the contrary, $\sin2\beta$ is measured with
precision and it is also more sensitive than $\gamma$. Thence, we have decided
to reconstruct the CKM mass matrix using $|V_{us}|$, $|V_{ub}|$ and $|V_{cb}|$,
while for the CP parameter we have chosen either $\sin2\beta$ or $\sin\gamma$.
Note that these two equivalent sets fully parametrize the CKM unitary matrix. In
what follows, we assume the elements $V_{ud}$, $V_{us}$, $V_{cb}$ and $V_{tb}$
positive, without loss of generality.

To fully reconstruct the CKM matrix from the input quadruplet $\left(V_{us},\,
V_{ub},\,V_{cb},\,\sin2\beta\right)$ we make use of the Branco-Lavoura (BL)
parametrization~\cite{Branco:1988ba}, which depends on $\lambda$, $A$, $\mu$
and the CP phase $\phi$. On the other hand when we reconstruct the
CKM matrix from $\left(V_{us},\, V_{ub},\,V_{cb},\,\sin\gamma\right)$ we use the
standard parametrization (SP)~\cite{Chau:1984fp}, which is defined by three
mixing angles $\theta_{12}$, $\theta_{13}$, $\theta_{23}$ and the
Kobayashi-Maskawa phase $\delta$. As we have mentioned in
Section~\ref{sec:numerics} for both reconstructions we ensure that the value of
the CP violating parameter $J$ is within the experimental range given
in Eq.~(\ref{eq:Jnum}).

\subsection{Reconstructing CKM with $\left(V_{us},\, V_{ub},\,
V_{cb},\,\sin2\beta\right)$}

In order to reconstruct the CKM mixing matrix, $V$, as a function of $V_{us},\,
V_{ub},\,V_{cb}$ and $\sin2\beta$ it is useful to use the BL parametrization,
which is a Wolfenstein-type parametrization, where the parameters $\lambda$, $A$
and $\mu$ are given by
\begin{equation}
\lambda=V_{us}\,,\quad
A=V_{cb}/V_{us}^2\,,\quad
\mu=\frac{|V_{ub}|}{V_{us}V_{cb}}\,,
\end{equation}
and the CP phase is given by $\phi=-\arg(V_{ub})$.
Once the quantities $V_{us}$, $V_{ub}$, $V_{cb}$ are given, the parameters
$\lambda$, $A$, $\mu$ are determined. The CP phase $\phi$ is related to the
$\sin2\beta$ as,
\begin{equation}
\sin\phi=-\frac{U_{cd}U_{tb}U_{cb}U_{td}}{2V_{us}V_{cs}V_{cb}|V_{ub
}| Q}\,\sin2\beta\,,
\label{eq:sph}
\end{equation}
where $U_{ij}\equiv|V_{ij}|^2$. The elements $U_{cd}$,
$U_{td}$ and $U_{tb}$ are determined by invoking the unitary conditions of
$V$
\begin{align}
U_{cd}&=\,1-U_{cs}-U_{cb}\,,\\
U_{tb}&=\,1-U_{cb}-U_{ub}\,,\\
U_{td}&=\,U_{cs}+U_{cb}+U_{us}+U_{ub}-1\,.
\end{align}
The quantity $Q$ in Eq.~(\ref{eq:sph}) is then given by
\begin{equation}
Q=\frac12\left(1-U_{cd}-U_{tb}-U_{cb}-U_{td}+U_{cd}
U_{tb}+U_{cb} U_{td}\right)\,.
\end{equation}
Finally, $V_{cs}$ is written as
\begin{equation}
\begin{split}
V_{cs}=&\left[-(U_{us}U_{cb}U_{ub})^{1/2}\cos\phi+(\,1-U_{us}-U_{cb}
\right.\\
&+U_{us}U_{cb} -2U_{ub}+U_{us}U_{ub}+U_{cb}U_{ub}\\
&\left.+U_{ub}^2 -U_{us}U_{cb}U_{ub}\sin^2\phi\,)^{1/2}
\right]/(1-U_{ub})\,.
\end{split}
\label{eq:vcs}
\end{equation}
Therefore, by solving Eq.~(\ref{eq:sph}) we determine $\sin\phi$.

\subsection{Reconstructing CKM with $\left(V_{us},\,V_{ub},\,V_{cb},\,
\sin\gamma\right)$}

To reconstruct the CKM mixing matrix using $V_{us}$, $V_{ub}$, $V_{cb}$ and
$\sin\gamma$ we take advantage of the SP,
where the CKM matrix is given as a function of the angles $\theta_{12}$,
$\theta_{13}$ and $\theta_{23}$ through the relations
\begin{align}
\sin\theta_{12}&=\,\frac{V_{us}}{\sqrt{1-|V_{ub}|^2}}\,,\\[2mm]
\sin\theta_{13}&=\,|V_{ub}|\,,\\[1mm]
\sin\theta_{23}&=\,\frac{V_{cb}}{\sqrt{
1-|V_{ub}|^2}}\,.
\end{align}
The CP violating phase $\delta$ can be
determined from the input parameters $V_{us}$, $V_{cb}$,
$|V_{ub}|$ and the angle $\gamma$
by solving the equation
\begin{equation}
\sin^2\delta=\left[1+
\tfrac{2\left|V_{ub}\right|V_{cb}V_{ud}}{V_{us}V_{tb}}\cos\delta+
\left(\tfrac{\left|V_{ub}\right|V_{cb}V_{ud}}{V_{us}V_{tb}}\right)^2
\right]\,\sin^2\gamma\,,
\label{eq:gammaSP}
\end{equation}
where $V_{ud}=\sqrt{1-V_{us}^2-|V_{ub}|^2}$. We remark that the phases
$\phi$ and $\delta$ from the BL and SP parametrization can
be related using the fact that the CKM matrix element $V_{cs}$ is real in
the BL parametrization and complex in SP,
\begin{equation}
\tan\phi=\cfrac{\tan\delta}{1-\cfrac{V_{cb}V_{us}|V_{ub}|}{
V_{tb}V_{ud}\cos\delta}}\,.
\label{eq:phases}
\end{equation}
Taking into account the hierarchy of $V$ we have verified that
$\tan\phi\simeq\tan\delta$ is a good approximation.

\end{document}